# NAMED ENTITY RECOGNITION USING WEB DOCUMENT CORPUS

WAHIBA BEN ABDESSALEM KARAA

*Institut Superieur de Gestion de Tunis 41 avenue de la Liberté*

Cité Bouchoucha, Le Bardo 2000, Tunisia,
wahiba.abdessalem@isg.rnu.tn

*Abstract*

This paper introduces a named entity recognition approach in textual corpus. This Named Entity (NE) can be a named: location, person, organization, date, time, etc., characterized by instances. A NE is found in texts accompanied by contexts: words that are left or right of the NE. The work mainly aims at identifying contexts inducing the NE's nature. As such, The occurrence of the word "President" in a text, means that this word or context may be followed by the name of a president as President "Obama". Likewise, a word preceded by the string "footballer" induces that this is the name of a footballer. NE recognition may be viewed as a classification method, where every word is assigned to a NE class, regarding the context.

The aim of this study is then to identify and classify the contexts that are most relevant to recognize a NE, those which are frequently found with the NE. A learning approach using training corpus: web documents, constructed from learning examples is then suggested. Frequency representations and modified *tf-idf* representations are used to calculate the context weights associated to context frequency, learning example frequency, and document frequency in the corpus.

## 1. Introduction

Named Entity Recognition is complex in various areas of automatic Natural Language Processing of (NLP), document indexing, document annotation, translation, etc. [2]. It is a fundamental step in various Information Extraction (IE) tasks. It has been an essential task in several research teams such as the Message Understanding Conferences (MUC), the Conferences on Natural Language Learning (CoNLL), etc.

Named entities (NE) are phrases containing the names of persons, locations, etc. They are particularly important for the access to document content, since they form the building blocks upon which the analysis of documents is based.

This paper discusses the use of learning approach for the problem of NE recognition. The goal is to reveal contextual NE in a document corpus. A context considers words surrounding the NE in the sentence in which it appears, it is a sequence of words, that are left or right of the NE. We use, in this work, the makings of learning technologies, combined with statistical models [17], to extract contexts from Web document corpus, to identify the most pertinent contexts for the recognition of a NE. We investigate the impact of using different feature weighting measures, in the hope that they will yield more context classification.

The remaining of the paper is organized as follows: section 2, introduces the state of the art of methods applied in Named Entity Recognition. Section 3 describes the methodology and section 4 gives test results of our approach. Section 5 is devoted the work's conclusion.

## 2. State of the art

Named entity recognition can be used to perform numerous processing tasks in various areas: Information extraction systems [6], text mining [8], [16], Automatic Speech Recognition (ASR) [5], etc.

Several works are particularly interested in the recognition of named entities. Mikheev et al. [12] have built a system for recognizing named entities, which combines a model based on grammar rules, and statistical models, without resorting to named entity lists.

Collins et al. [1] suggests an algorithm for named entity classification, based on the meaning word disambiguation, and exploits the redundancy in the contextual characteristics.

This system operates a large corpus to produce a generic list of proper nouns. The names are collected by searching for a syntax diagram with specific properties. For example, a proper name is a sequence of consecutive words in a nominal phrase, etc.

Petasis et al. [14] presented a method that helps to build a rules-based system for recognition and classification of named entities. They have used machine learning, to monitor system performance and avoid manual marking.

In his paper [10], Mann explores the idea of fine-grained proper noun ontology and its use in question answering. The ontology is built from unrestricted text using simple textual co-occurrence patterns. This ontology is therefore used on a question answering task to provide primary results on the utility of this information. However, this method has a low coverage.

The Nemesis system presented by Fourour [6] is founded on some heuristics, allowing the identification of named entities, and their classification by detecting the boundaries of the entity called "context" to the left or right, and by studying syntactic, or morphological of these entities. For example, acronyms are named entities consisting of a single lexical unit comprising several capital letters, etc.

Krstev et al [9] suggested a basic structure of a relational model of a multilingual dictionary of proper names based on four-level ontology. However, the implementation is not yet completed, it is only expected.

The KNOWITALL system planned by Etzioni et al [4] aims at automating the process of extracting named entities from the Web in an unsupervised and scalable manner. This system is not intended for recognizing a named entity, but used to create long lists of named entities. However, it is not designed to resolve the ambiguity in some documents.

Friburger [7] recommends a method based on rules for finding a large proportion of person names. However, this method has some limitations as errors, and missing responses.

Nadeau et al. [13] have suggested a system for recognizing named entities. Their work is based on those of Collins et al. [1], and Etzioni et al. [4]. The system exploits human-generated HTML markup in Web pages to generate gazetteers, then it uses simple heuristics for the entity disambiguation in the context of a given document.

Marthineau et al. [11] showed the validity of existing local grammars in the system GRAALWEB to recognize and extract the named entities.

In view of works touching the recognition of named entities, we perceive that most of them are based on a set of rules in relation to predefined categories: morphological, grammatical, etc. [7], [11], [14], or on predefined lists of dictionaries [9]. The ontology domain is still in exploration [10].

We adopted the idea of Nemesis based on the left and the right context of the named entity. However, our approach does not mark the context derived from syntactic or morphological rules, but identifies the context founded on learning phase. The objective is thus to carry out a system, able to induce the nature of a named entity, following the meeting of certain indicators, and this, in any language, without requiring to dictionaries or lists of named entities.

For our approach we use, for the learning phase, documents resulting from the Web. We use context frequency representations, based on *tf-idf* version, to calculate context weights. These weights are used to determine the most pertinent contexts, to identify the named entity nature.

## 3. Methodology

The system encloses three main phases. The first is the training corpus collection. The second is the context extraction and the contexts classification, according to weighting measures. The third phase is the weighting measures exploitation, in order to build a model for NE recognition. The figure (Fig 1) summarizes the architecture.

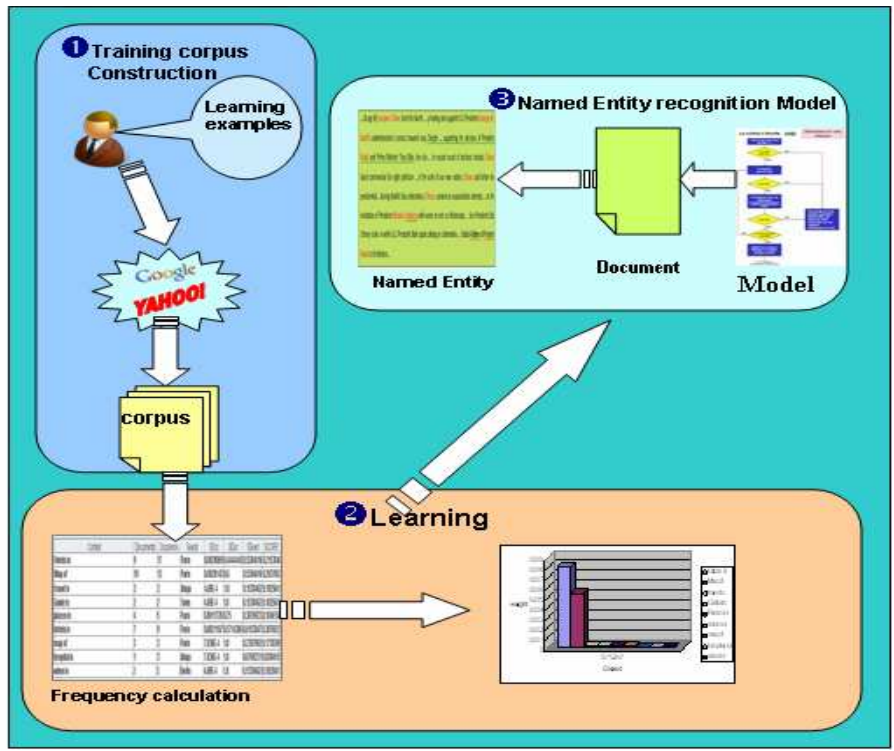

Fig. 1. The architeture

### 3.1. *Training corpus construction*

The first step is to build an initial corpus containing web documents. This corpus is called learning corpus. Our algorithm proceeds as follows:

(i) We Furnish a set of learning examples, which are instances of a named entity class. For example for the class entity "disease", we provide as learning examples, names of diseases such as "flu", "Eczema", "measles", etc. For the named entity class "football player" must be provided as instances, footballer names such as "platini", "Zidane", "Maradona", etc.

(ii) For all the instances a query is sent to a Web search engine (Yahoo, Google, AltaVista, etc.), by using a specific API related to each engine, searching for context string surrounding the instances: words before, or/and after the instance, the length of the string is an input parameter.

(iii) The result is a list of links. Afterwards, for each link, a new request is released in order to retrieve and save the document. To increase the number of returned links, the user should provide a maximum of learning examples. The figure (Fig 2) is an extract of the API used with yahoo engine:

```
import com.yahoo.search.SearchClient;
import com.yahoo.search.SearchException;
import com.yahoo.search.WebSearchRequest;
import com.yahoo.search.WebSearchResult;
import com.yahoo.search.WebSearchResults;
public class Extract_Req_Thread extends Thread {
. . .
SearchClient client = new SearchClient("Yahoo key");
WebSearchRequest request = new WebSearchRequest("instance of the named entity");
WebSearchResults results = client.webSearch(request);
. . .
for (int i = 0; i < results.listResults().length; i++) {
WebSearchResult result = results.listResults()[i];
//Traitements
}
```

Fig. 2. The API algorithm

### 3.2. *Learning phase*

A context is a set of words preceding, or following a named entity. We focus, in this study, on the words that precede a named entity instance; we consider that the left context is more relevant then the right one.

For instance, we suggest the concept "President" with the six instances, considered as learning examples: "Bush", "G W. Bush", "Nicolas Sarkozy," Sarkozy "," Chirac ", and "Jacques Chirac". The result of the extraction phase is a set of context candidates. The figure (Fig. 3.) is an extract of a document where the context candidates are pointed:

> …At **age 69**, <u>Jacques Chirac</u> faced fourth…a leading voice against U.S. **President** <u>George W. Bush</u>'s administration's conduct towards Iraq. Despite … supporting the decisions of **President** <u>Bush</u> and Prime Minister Tony Blair. See also …the second round of elections, **instead**, <u>Chirac</u> faced controversial far right politician … of the unity of our **nation,** <u>Chirac</u> said before the presidential…during Bastille Day **celebration,** <u>Chirac</u> survived an association attempt … at the invitation of **president** <u>Nicolas Sarkosy</u> with whom he met on Wednesday … Vice president Dick Cheney looks on while U.S. President Dick Cheney looks on while U.S. **President** <u>Bush</u> during an alternative …

Fig. 3. Example of extracted document

There are eight instances with different contexts. We can notice that the context "President" is most common with the learning examples, which makes it the more appropriate to recognize the name of a president. However, the context occurrence number is an insufficient parameter to decide that it is the best context to construct a rule:

**President** *<president-name>*.

The goal is to find high-quality context. To define the context's quality, we observe some metrics calculated through frequencies, given that instances found on the Web surrounded by the context.

Estimating pertinence of contexts is not sufficient by only the Frequency of these contexts found surrounding a NE instance, especially when we have no labelled negative learning examples, but only positive ones.

The purpose of this work is to identify the most pertinent contexts for the identification of a given named entity. For this intention we calculate weights to classify contexts taking into account the context frequency regarding learning examples, the inverse context frequenc*y,* the learning example frequency in the document corpus, and the frequency of documents containing contexts in the document corpus. To calculate context weights, we use context frequency representations based on *tf-idf* version. *tfidf* (Term Frequency-Inverse Document Frequency) is one of the most classics and most common weighting method used to describe documents in the Vector Space Model [17], particularly in Information Retrieval (IR). The *tf* considers the term frequency in the document: the more a word occurs in a document, the more it is expected to be significant in this document. In addition, *idf* inverse document frequency measures the term frequency in the corpus: the more a word appears in a corpus, the more it is estimated irrelevant for the document.

The Term Trequency *(tf )* of a term *ti* for a document *dj* is calculated as follows:

$$tf_{ij} = \frac{frequency_{ij}}{\sum_k frequency_{kj}} \quad (1)$$

Where *frequency $_{kj}$* is the occurrence number of the term *ti* in *dj*. The denominator is the number of occurrences of all terms in the document *dj*.

The *idf*, Inverse Document Frequency component is computed as follows:

$$idf_i = \log \frac{N}{n_i} \quad (2)$$

Where *N* is the total number of documents in the corpus, and *ni* is the number of documents in which the term *ti* emerges. In *tf.idf* weighting is:

$$w_{ij} = tf_{ij} \times idf_i \quad (3)$$

Many alternatives have been suggested to the basic *tfidf* formula, where the *tf* or *idf* part is modified using functions related to characteristic selection. It is in our case to measure the importance of a context *ci* to a NE; in our work we propose a modified formula. We define:

**Definition 1:** *context frequency*

The context frequency *(cfi)* of a context *ci* in a document corpus is calculated as follows:

$$cf_i = \frac{nc_i}{\sum nc_i} \quad (4)$$

We consider the variable *nci* the occurrence number of a context within a document corpus accompanied by a NE Learning Example (LE), and $\sum nc_i$ is the occurrence number of all contexts in the document corpus.

**Definition 2:** *learning example frequency*

$$lef_i = \frac{nle_i}{NLE} \quad (5)$$

We designated *nlei* the number of LE located with the context *ci,* in the corpus, and *NLE* the total number of learning examples used for training.

**Definition 3:** *document frequency*

We used the variable *ndi* to assign the occurrence number of documents in the corpus containing a context *ci*, and coming from different sources. *Di* is the occurrence number of documents in the corpus containing a context *ci*.

The document frequency *dfi* is calculated as follows:

$$df_i = \frac{nd_i}{D_i} \quad (6)$$

**Definition 4:** *inverse context frequency*

We considered the hypothesis that the context is significant for a NE, if it does not often appears with other phrases in the corpus. *icf* inverse context frequency, measures the context frequency associated to all the phrases escorted by the context: the more a context appears in a corpus accompanied with other phrases, the more it is estimated irrelevant for the NE.

*icfi* is calculated using *nci*, the occurrence number of a context within a document corpus accompanied by learning example, and the variable *Ci*, which represents the total occurrence number of the context accompanied with other phrases in the corpus.

$$icf_i = \frac{nc_i}{C_i} \quad (7)$$

**Definition 5:** *context weight*

The product of the obtained frequencies provides the weight *wi* adapted to the detection of the context pertinence:

$$w_i = cf_i \times lef_i \times df_i \times icf_i \quad (8)$$

### 3.3. *Recognition model*

NE recognition may be viewed as a classification problem, where every word is assigned to a NE class, regarding the context. Since decision tree algorithms are widely used for data mining [3], classification, etc. in this third phase, we design algorithm recognition, in the form of a decision tree, based on C4.5 algorithm [15].

In many cases, the same context can introduce different NE. For example, the context "Mr." precedes both a footballer name and a president name: Mr. Zidane, Mr. Obama. The intention is to decide whether the NE is a president name or a footballer name.

We define $vote_i$ for each NE with each $context_i$ in the document. The value of $vote_i$ is incremented with the weight of the context, each time a $context_i$ is encountered:

$$vote_i = vote_i + w_i \qquad (9)$$

Afterward, the algorithm decides the nature of the NE, once the value of $vote_i$ reaches a threshold, and depending on the value of $vote_j, \ldots, vote_n$ reached by the other contexts.

## 4. Experimental results

We conducted several tests with different NE. For example for the NE class "capital", we extracted a corpus of documents from 65 URLs, obtained by queries on the search engine Yahoo. For the 13 following instances (or learning examples): *"Paris", "Tunis", "Cairo", "Athens", "Abuja", "Berlin", "Bucharest", "Budapest", "Brasilia", "Freetown", "Dublin", "Vienna", and "Doha"*. We obtained 2398 contexts composed of two different words to the left. The learning examples are occurred 4264 times in the corpus with the contexts. We present in the following table an extract of the obtained context classification results:

Table 1. Context classification.

| Context | cf | df | lef | icf | W |
|---|---|---|---|---|---|
| Hotels in | 0,0039869 | 0,4444445 | 0,5384616 | 8,5 | 0,008110106 |
| Map of | 0,0028143 | 0,6 | 0,5384616 | 6 | 0,005455413 |
| travel to | 0,000469 | 1 | 0,1538462 | 2 | 0,000144308 |
| Guide to | 0,000469 | 1 | 0,1538462 | 2 | 0,000144308 |
| Places in | 0,0011726 | 0,75 | 0,3076923 | 1 | 0,0002706 |
| hotels in | 0,0021107 | 0,5714286 | 0,6153847 | 3 | 0,002226673 |
| map of | 0,0007036 | 1 | 0,2307692 | 1 | 0,000162369 |
| Hospital in | 0,0007036 | 1 | 0,0769231 | 0,4 | 2,16E-05 |
| when in | 0,000469 | 1 | 0,1538462 | 0,4 | 2,89E-05 |

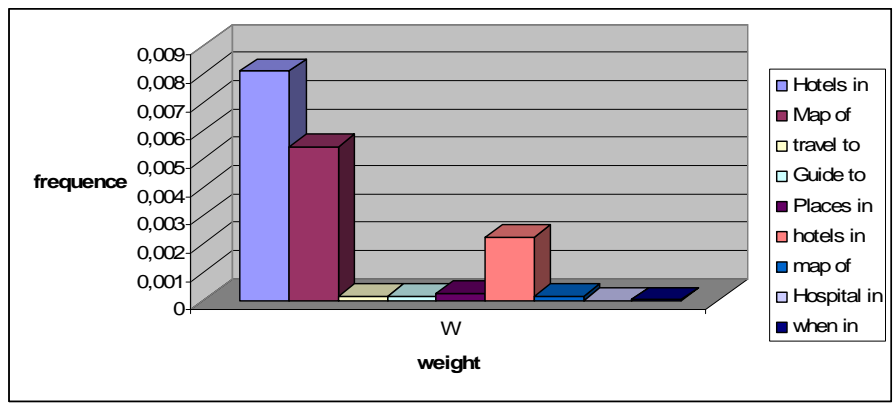

Fig. 4. Context weight classification

In the table (Table 1), we note that the context "Hotel in" is found 17 times, in the corpus with the learning examples. The context frequency is:

$$cf_{Hotel\ in} = \frac{17}{4264} = 0,0039869 \qquad (10)$$

The context is found in 9 documents, but among these documents, only 4 documents come from different sources. The document frequency for this context is:

$$df_{Hotel\ in} = \frac{4}{9} = 0,4444445 \quad (11)$$

The context "Hotel in" occurs 17 times in total, but only with 7 learning examples, the learning example frequency is then:

$$lef_{Hotel\ in} = \frac{7}{13} = 0,5384616 \quad (12)$$

The context occurs 17 times with learning examples, but also 2 times with other phrases, the inverse context frequency is:

$$icf_{Hotel\ in} = \frac{17}{2} = 8,5 \quad (13)$$

The weight of this context is then:

$$W_{Hotel\ in} = cf_{Hotel\ in} \times df_{Hotel\ in} \times lef_{Hotel\ in} \times icf_{Hotel\ in} = 0.008110106 \quad (14)$$

The context "Hotel in" has the highest weight, which makes it more relevant for the recognition of a capital name, compared to the other contexts.

We studied in addition, if the context number may change if we increase the size of the analyzed corpus (the document number). In the table (Table 2), we present some statistics associated to the ratio between the contexts number, and the corpus size. These numbers are found for the NE class "President", using 89 representations of president names as learning examples:

"Jacques Chirac", "JACQUES CHIRAC", "Chirac", "Chirac", "CHIRAC", "Jacques CHIRAC"; "Bush", "George Walker Bush", "George Bush", "GEORGE BUSH", "GEORGE W BUSH", "GEORGE W. BUSH", "George W Bush", "BUSH", "George W. BUSH", "Geroge BUSH", "Georges W.", "Georges W", "Nicolas Sarkozy", "NICOLAS SARKOZY", "Sarkozy", "Sarko", "sarkozy", "Nicolas sarkozy", "Nicolas SARKOZY", "SARKOZY", "Jalal Talabani", "Talabani", "TALABANI", "Jalal TALABANI", "JALAL TALABANI", "Zine El Abidine Ben Ali", "Zine el Abidine", etc.

Table 2. Corpus size influence.

| Document number | Learning Example occurrence number | Context number (2 words left) |
|---|---|---|
| 80 | 3003 | 1911 |
| 110 | 3637 | 2379 |
| 160 | 4699 | 2850 |
| 200 | 6894 | 3120 |
| 240 | 8592 | 3215 |
| 280 | 9260 | 3496 |
| 320 | 10362 | 3516 |
| 360 | 11184 | 3578 |
| 420 | 12218 | 3618 |
| 460 | 12981 | 3634 |

We have noticed that the context number becomes stable from a significant size corpus (420 records = 7 MB of text data).

Recall and precision are usually admitted approachs of measuring system performance in this field. We suppose that Recall is the number of correct NE found by the system over total number of correct NE in the document corpus. Precision is the number of correct NE found by the system over total number of NE found in the corpus. We obtained the following value: for Precision 80,9 % and recall 69,8 %.

## 5. Conclusion

In this paper, the use of learning approach is suggested for the problem of NE recognition. The goal is to uncover in a document corpus, NE that occurs frequently accompanied by contexts: sequence of words, that are left or right of the NE. Contexts that occur with given learning examples were first extracted from Web documents corpus. Different feature weighting measures were examined to classify the contexts in order to identify the most pertinent contexts for the recognition of a NE. This classification enables to derive a model for the recognition of a NE.

The same strategy can be applied to person names, company names, and many other types of named entities in any language. Although, we should mention that we successfully applied this technique to several named-entity types, in French and in English language.

One of the future works which we recommend is to discern and to measure similarity between contexts. We can use this measurement to cluster similar contexts.

Since we have primarily applied our approach to the named entities problem, we can also attempt additional concepts. In effect, the context usually contains enough information to identify the instance as a concept member.

## References

[1] M. Collins and Y. Singer, Unsupervised models for named entity classification, *in Proceedings of the Joint SIGDAT Conference on Empirical Methods in Natural Language Processing and Very Large Corpora*, 1999, pp. 189–196,

[2] B. Daille, and E. Morin, Reconnaissance automatique des noms propres de la langue écrite: les récentes TAL, *Traitement automatique des langues*, 2000, vol. 41, n° 3, pp. 601-621.

[3] F. Denis, R. Gilleron, and F. Letouzey, Learning from positive and unlabeled examples. *Elsevier. Theoretical Computer Science,* 2005, vol. 348, pp. 70 – 83

[4] O. Etzioni, M. Cafarella, D. Downey, S. Kok, A. Popescu, T. Shaked, S. Soderland, D. Weld, and A.Yates, Unsupervised named-entity extraction from the web: An experimental study, *Artificial Intelligence*, 2005, vol. 65, pp. 91–134.

[5] B. Favre, F. Béchet, and P. Nocéra, Robust Named Entity Extraction from Spoken Archives, *in Proceedings of HLT-EMNLP'05*, pp. 491-498, Vancouver, Canada, October 2005.

[6] N. Fourour, and E.Morin, Apport du Web dans la reconnaissance des entités nommées. *Revue québécoise de linguistique,* 2003, vol. 32, n° 1, pp. 41-60.

[7] N. Friburger, Linguistique et reconnaissance automatique des noms propres, *Meta : journal des traducteurs,* 2006, vol. 51, n° 4, pp. 637-650.

[8] C. Jacquemin, and C. Bush, Fouille du Web pour la collecte d'entités nommées*, in Proceedings 8ème Conférence Nationale sur le Traitement Automatique des Langues Naturelles (TALN 2000)*, pp. 187-196. Lausanne, 2000.

[9] C. Krstev, D. Vitas, D. Maurel, M. Tran, Multilingual ontology of proper name, *in Proceedings of the Language and Technology Conference,* pp. 116–119, Poznan, Poland, 2005.

[10] G.S. Mann, Fine-grained proper noun anthologies for question answering, *International Conference on Computational Linguistics*, COLING-02 on SEMANET: building and using semantic networks, 2002, Vol. 11, pp. 1-7.

[11] C.Martineau, E. Tolone, and S. Voyatzi, Les Entités Nommées : usage et degrés de précision et de désambiguïsation. *In proceedings of 26th conference on Lexis and Grammar,* pp. 105-112, Bonifacio, 2-6 October 2007.

[12] A. Mikheev, M. Moens, and C. Grover, Named Entity Recognition without Gazetteers, *in Proceedings of Conference of European, Chapter of the Association for Computational Linguistics, EACL '99,* pp. 1-8, University of Bergen, Bergen, Norway June 1999.

[13] D. Nadeau, P. D. Turney, and S. Matwin, Unsupervised named-entity recognition: Generating gazetteers and resolving ambiguity, *Lecture Notes in Computer Science, Springer,* 2006, pp. 266–277, Berlin Heidelberg 2006.

[14] G Petasis, F Vichot, F Wolinski, G Paliouras, V. Karkaletsis, and C. D. Spyropoulos, Using machine learning to maintain rule-based named-entity recognition and classification, *in Proceedings of the 39th Annual Meeting on Association for Computational Linguistic*s, pp. 426 – 43, Toulouse, France, 2001.

[15] J.R. Quinlan, C4.5: Programs for Machine Learning, Morgan Kaufmann, Los Altos, CA, 1993.

[16] S. K. Saha, S. Sarkar, and P. Mitra, Feature selection techniques for maximum entropy based biomedical named entity recognition, *Journal of Biomedical Informatics*, Vol. 42, Issue 5, pp. 905-911, Elsevier, October 2009.

[17] G. Salton, A. Wong,, and C.S. Yang, A vector space model for information retrieval, *Journal of the American Society for Information Science,* 1975; vol. 8(11), pp. 613-620.